\documentclass[12pt,a4paper]{article}

\usepackage{amsmath,amssymb, amsfonts, amsthm}

\title{Classical signal model reproducing quantum probabilities for single and coincidence detections}

\author{Andrei Khrennikov, B\"orje Nilsson, and Sven Nordebo\\
International Center for Mathematical Modelling
\\in Physics and Cognitive Sciences\\
Linnaeus University,  V\"axj\"o, S-35195, Sweden}

\begin{document}

\maketitle

\begin{abstract} 
We present a simple classical (random) signal model reproducing Born's rule.   
The crucial point of our approach is that the presence of  detector's threshold and calibration 
procedure have to be treated not as simply experimental technicalities, but as the basic counterparts of the theoretical model. We call this approach  threshold signal detection model (TSD). 
The experiment on coincidence detection which was done by Grangier in 1986 \cite{Grangier} played 
a crucial role in rejection of (semi-)classical field models in favor of
quantum mechanics (QM): impossibility to resolve the wave-particle duality in favor of a purely wave model.
QM predicts that the relative probability of coincidence detection, the coefficient $g^{(2)}(0),$ 
is zero (for one photon states), but in (semi-)classical models $g^{(2)}(0)\geq 1.$ 
In TSD the coefficient $g^{(2)}(0)$  decreases as $1/{\cal E}_d^2,$ where ${\cal E}_d>0$ 
is the detection threshold. Hence, by increasing this threshold an experimenter can make the coefficient $g^{(2)}(0)$ essentially less than 1. The TSD-prediction can be tested experimentally in new Grangier type experiments presenting a detailed monitoring of dependence of the coefficient $g^{(2)}(0)$ on the detection threshold. 
\end{abstract}

\section{Introduction}

We study the old problem of a possibility to construct a classical field model reproducing probabilistic predictions of quantum mechanics. The common opinion, see Bell  \cite{B},  is that this is impossible to do, but see \cite{DBR}--\cite{Theo} for numerous attempts to proceed with classical wave (oscillatory) prequantum models. The main argument is that for composite quantum systems (e.g., a pair of entangled photons) the correlation predicted by QM (and confirmed by experiment \cite{ASP}--\cite{WEIHS1}) cannot be reproduced by models with (local) hidden variables. This is also the common opinion that ``Bell's theorem''  \cite{B} is formulated in so general abstract framework that it rejects all (local) models with hidden variables, including variables of classical field (e.g.,
electromagnetic) type.  Although nowadays Bell's inequality  and entangled systems are the hot topics 
\cite{KV09}, \cite{KV08},
the problem of a possibility of the classical field description of a single quantum system is also of the large
importance for quantum theory, including quantum information theory. One of the most important tests of a possibility
to represent a photon simply as a pulse of classical electromagnetic field is the experiment on coincidence detection 
in two output channels of the polarization beam splitter (PBS)  \cite{Grangier}, \cite{Grangier1}. QM predicts that if the source can be considered as 
one-photon source, then the probability of coincidence detection equals to zero: one photon  cannot be split between two
channels, it is either in one or another channel. Any pulse of the classical electromagnetic field is split between two 
output channels of PBS. Hence, the probability of coincidence detection is nonzero. Of course, the real experimental situation
is more complicated: for any ``one-photon source'' the probability
 of emission of e.g. two photons is nonzero (although it can be made very small), 
there is also the contribution of noise, including so called duck counts (i.e., counts in detectors
in the absence of the source). Therefore, instead of the absolute probability of coincidence detection   $P_{12},$
experimenters use the relative probability  \cite{Grangier}, \cite{Grangier1}:
\begin{equation}
\label{CHCH1_9}
g^{(2)}(0)= \frac{P_{12}}{P_1 P_2},
\end{equation} 
where $P_{12}$ is the probability of coincidence detection in channels $i=1,2$ and $P_i$ are probabilities of 
detection in corresponding channels. Known (semi-) classical models of QM predict that $g^{(2)}(0)\geq 1.$ 
If photon is not a classical pulse, one can expect that 
$g^{(2)}(0) <1$ (even by taking into account noise and emission of double photons).  The first experiment of this type was performed by \cite{Grangier}, see also 
\cite{Grangier1} and \cite{Beck} for review. It was shown that the number of double clicks is relatively small. 
 This experiment played an important role in quantum foundations. As is commonly accepted, this experiment justified rejection of (semi-)classical field theories as ``prequantum theories''; in particular, photons cannot be interpreted as pulses of classical electromagnetic field. 

The aim of this paper is to show that the main reason of dis-matching of predictions of (semi-)classical field models with 
QM and experimental data is that such models do not take into account the impact of {\it detectors} to creation of 
the quantum statistics. 

We present a classical field model which in combination with the procedure of detection of random signals 
by {\it threshold type detectors} (which are properly calibrated) reproduces quantum probabilities and, 
in particular, the coefficient $g^{(2)}(0)<<1$ for a sufficiently high detection threshold ${\cal E}_d.$  
Hence, the prediction of our model differs crucially 
from the prediction of the known classical field  models which do not take into account the evident fact 
that measurement is not simply detection (of the monitoring-type) of continuous classical signals, 
but creation of discrete counts with the aid of threshold detectors.
Detectors also have to be properly calibrated. 

Our model, {\it threshold signal detection model} (TSD), predicts that the quantum prediction, $g^{(2)}(0)<<1,$ matches better measurements for high value of the detection threshold. 
The basic prediction of TSD is that the coefficient 
$$
g^{(2)}(0)\leq \frac{K}{{\cal E}_d^2},
$$
where $K>0$ is a constant depending on the signal.
This $K$ depends on the brightness of the source: higher brightness implies larger 
$K.$  It also depends on elements of the density matrix $\rho=(\rho_{ij})$ 
corresponding to the prequantum signal. (In TSD $\rho$ corresponds to the normalization (by the trace) 
of the covariance operator $B$ of the prequantum random signal, i.e., $\rho= B/\rm{Tr}B.)$
Larger $\rho_{12}$ implies larger  $K.$ 

It is interesting to compare this prediction with the real experiment. Unfortunately, it seems 
that Grangier's type experiments with detailed monitoring of dependence  of the coincidence probability 
on the value of the threshold have never been done;  more specifically, that is: In Grangier's
 experiment, did the calibration of detectors play a crucial role to
 eliminate coincidences?

 This is clearly a crucial question. It is hard to give a definite
 answer as to how exactly it was influencing the result at a fundamental
 level because Grangier et al \cite{Grangier}, \cite{Grangier1} didn't study the influence of threshold on the
 $g^{(2)}(0)$ parameter he was measuring. It is nevertheless clear that with a
 lower threshold, he would have gotten a less good result.
Let me translate the part of Grangier's thesis \cite{Grangier1} where he explains for
 the first time the role of the  threshold and how its
 level was chosen:
 
``[...] In this configuration, the threshold has a double role of
 acquisition of timing information and of selection of the pulses (the
 too weak pulses are not taken into account).
     The problems connected to the choice of the discriminator threshold
 and of the high voltage of the photomultipliers are discussed in detail
 in reference \cite{ASP}. We have in the present experiment chosen a {\bf rather
  high threshold}, which amount to give the priority of the stability of
 the counting rates and the reproducibility of the results, rather than
 to the global detection efficiencies.'' (I stressed with bold the important 
 fact that Grangier proceeded with rather high threshold.)

{\it TSD provides a strong motivation to perform Grangier's type experiment with monitoring of dependence of the coincidence 
probability on the detection threshold and source's brightness.}

TSD can be considered as measurement theory for recently developed {\it prequantum classical statistical field theory}, PCSFT, \cite{KH1},\cite{W1}.  The latter reproduced all quantum averages and correlations including correlations for entangled quantum states. In particular,  PCSFT correlations violated Bell's inequality. 
The main problem for matching of PCSFT and conventional QM was that PCSFT (nor other classical field models) was not able to describe
probabilities of discrete clicks of detectors. In particular, PCSFT  is  theory of correlations of {\it continuous signals}.
``Prequantum observables''  are given by quadratic forms of signals. These forms are unbounded and this is not surprising that correlations of such observables can violate Bell's type inequalities, see \cite{CONT} for discussion and an elementary example. The condition of coincidence of ranges of values of quantum observables and corresponding ``prequantum variables''
plays a crucial role in Bell's argument. TSD solved the measurement problem of PCSFT. In the same way as in Bell's consideration, TSD operates with discrete observables. In particular, in the case of photon polarization (its projection to a fixed axis) TSD operates with dichotomous variables taking values $\pm 1.$ 

TSD/PCSFT for composite quantum systems was presented in \cite{ARCH}. However, random signals considered 
in  \cite{ARCH} have a complex structure of temporal correlations, even in the case of a single system.
In the present paper, we use simply a combination of the Wiener process (as the career of temporal correlations) 
and a Gaussian stationary process valued in the space of internal degrees of freedom of a quantum system
(e.g., its polarization). We do not consider spatial degrees of freedom.  

To escape misunderstanding, we stress that the presented detection model is purely classical; in particular, detectors are classical 
detectors which are sensitive to the energy density of a signal in the domain of detection $V.$ Such a detector clicks at the first 
instance of time $\tau=\tau_{\rm{d}}$ when the total energy of signal in $V$ approaches the detection threshold ${\cal E}_d.$
Since we consider random signals, this instant of detection is a random variable $\tau=\tau(\omega).$ Our main aim is to find
its average $\bar{\tau}.$ The quantity $1/ \bar{\tau}$ determines the probability of detection.  At the moment we proceed with 
such an operational description of the classical detectors. We plan to consider a more detailed scheme of the classical
threshold detection in another paper in which we shall study classical signals with spatial degrees of freedom, i.e., 
signals in physical space-time. (In the present paper
we restrict the model to ``internal degrees of freedom'' such as polarization. The model with spatial degrees of freedom is essentially more
complicated, since it involves processes with infinite-dimensional state space.)

\section{Threshold detection}

\subsection{The class of random signals}

We consider a special class of classical random signals. The model is phemenological:
we cannot present physical motivations for selection of this class of signals, besides 
the fact that detection of such signals with the aid of threshold type detectors reproduces 
the correct quantum probabilities. Another class of classical random signals serving for the same aim 
was introduced in \cite{ARCH}. Signals considered in the present paper have essentially simpler
temporal structure, simply the Wiener process.

Structurally our model has some similarity with the prequantum model of Gr\"ossing  et al \cite{Grossing1}.
Subquantum stochasticity is combined of the two counterparts: a stationary process in the space of
internal degrees of freedom and the randmom walk type motion describing the temporal dynamics.

We start with stationary signals. We proceed in the finite dimensional state space corresponding 
to internal degrees of freedom such as polarization. Generalization to spatial degrees of freedom is evident,
but it has essentially more complicated mathematical structure.

Let $H$ be the $m$ dimensional complex Hilbert space. Let $\phi\equiv \phi(\omega)$ be the $H$-valued Gaussian random variable
with zero average and the covariance operator $B.$  This operator is Hermitian and positively defined. 
Take in $H$ an orthonormal basis $\{e_j\}$ and consider corresponding signal's components $\phi_j(\omega)=
\langle \phi(\omega), e_j\rangle.$ The mathematical expansion of the random vector $\phi(\omega)$ with respect 
to the basis $\phi(\omega)= \sum_j \phi_j(\omega) e_j$ physically corresponds to splitting of the signal into 
disjoint channels. We remark that correlations of signal's components are given by
$$
E \phi_i \overline{\phi_j}= \langle B e_i, e_j \rangle= b_{ij}.
$$
Now we introduce the temporal stochastics by simply using the one dimensional Wiener process $w(t)$
which is independent from the stationary process $\phi(\omega).$
Consider the random (non-stationary) signal $\phi(t, \omega)= w(t) \phi(\omega)$  and its components
corresponding to internal degrees of freedom: $\phi_j(t, \omega)= w(t) \phi_j(\omega).$
We remark that correlations of signal's components are given by
$$
E \phi_i(t) \overline{\phi_j(s)}= \min (t,s)  b_{ij}.
$$
The energy of the $i$th component of the complex signal $\phi(t)$ (at the instance of time $t)$ 
is given by the square of its absolute value
$$
{\cal E}_i (t, \omega)= \vert \phi_i(t, \omega)\vert^2.
$$
The total energy of the signal is
$$
{\cal E}_i (t, \omega)= \sum_i\vert \phi_i(t,\omega)\vert^2 =\Vert  \phi(t,\omega)\Vert^2.
$$

\subsection{The scheme of threshold measurement}

We consider the measurement scheme in which each channel, $i=1,2,...m,$ goes to a threshold type detector.
We assume that all detectors have the same threshold ${\cal E}_d >0.$
The detection procedure under consideration is reduced to 
the condition of the energy level approaching the detection threshold. The  instant of time $\tau$ corresponding to 
the signal's detection (``click'') by $j$th detector is determined by the condition:
\begin{equation}
\label{BE0a}
{\cal E}_j(\tau, \omega)=  {\cal E}_d.
\end{equation}
We remark that the instant of the signal detection is a random variable:
$$
\tau=\tau(\omega).
$$
Mathematically our aim is to find average of the instance of detection, $\bar{\tau}= E \tau.$ 
The quantity $1/\bar{\tau}$ will be used to find the probability of detection, ``how often the detector produces clicks,'' see section \ref{S2}.

We apply the mathematical expectation (average) operator to both sides of the detection condition 
(\ref{BE0a}) and we obtain 
\begin{equation}
\label{BE0a_Y}
E {\cal E}_j(\tau(\omega), \omega)=  {\cal E}_d,
\end{equation}
or 
\begin{equation}
\label{BE0a_Y1}
E w^2(\tau(\omega),\omega) E \vert \phi_j(\omega) \vert^2 =  {\cal E}_d.
\end{equation}
To find the first average, we use the formula of total probability:
$$
E w^2(\tau(\omega),\omega)  = \int_0^\infty d\tau P(\tau(\omega)=\tau) E w^2(\tau, \omega).
$$
We know that, for the fixed $\tau,$ $E w^2(\tau, \omega)= \tau.$ Hence,
 $$
E w^2(\tau(\omega),\omega)  = \int_0^\infty d\tau P(\tau(\omega)=\tau) \tau= E\tau\equiv \bar{\tau}.
$$
Thus, the detection condition (\ref{BE0a_Y1}) has the form:
\begin{equation}
\label{BE0a_Y2}
\bar{\tau} E \vert \phi_j(\omega) \vert^2 =  {\cal E}_d,
\end{equation}
or 
\begin{equation}
\label{BE0a_Y8}
\bar{\tau} b_{ii} =  {\cal E}_d.
\end{equation}
We remark that $\tau=\tau_i, i=1,2,..., m.$ Thus
\begin{equation}
\label{BE0c2X} 
 \frac{1}{\bar{\tau_i}}  =  \frac{b_{ii}}{ {\cal E}_d}.
\end{equation}

\section{Probabilities of clicks in detection channels}
\label{S2}

Hence, during a long period of time $T$ such a detector clicks $N_{\rm{click}}$-times, where
\begin{equation}
\label{BE0c3} 
 N_i \approx \frac{T }{\bar{\tau_i}} = \frac{b_{ii} T}{ {\cal E}_d}.
\end{equation}
To find the probability of detection and match the real detection scheme which is used in quantum experiments we have to use a proper normalization
of $N_i,$  This is an important point of our considerations. (The normalization problem is typically ignored in standard 
books on quantum foundations, cf., however,   \cite{CONT}.) In QM-experiments probabilities are obtained
through normalization corresponding to the sum of clicks in all detectors involved in the experiment, e.g., spin up and spin down  detectors. 

Hence, the total number of clicks:
 \begin{equation}
\label{BE0c5} 
N=\sum_i N_i =  \frac{T \sum_i b_{ii} }{{\cal E}_d}= \frac{T \rm{Tr} B }{{\cal E}_d}.
\end{equation}
We remark that the total number of clicks does not depend on the split of the signal into disjoint channels, i.e., 
on the selection of an orthonormal basis $\{e_j\}$ in $H.$ In fact, 
$$
E \Vert \phi(\omega)\Vert^2= \rm{Tr} B.
$$

The probability of detection for the $j$th detector is given by
\begin{equation}
\label{BE0c6} 
P_i= N_i/N = \frac{b_{ii}}{\rm{Tr} B}.
\end{equation}
In fact, this is the Born's rule of QM. Consider the operator
\begin{equation}
\label{BE0d2}
\rho= B/\rm{Tr} B.
\end{equation}
This is the Hermitian positive trace one operator; so  it has all properties of the {\it density operator} used in QM
to describe the state of a quantum system.

Set $\widehat{C}_i=\vert e_i\rangle \langle e_i\vert,$
the orthogonal projection onto the vector $e_i.$
 Then the equality for the probability of detection (\ref{BE0c6}) can be written as
\begin{equation}
\label{BE0d3}
P_i =\rm{Tr} \rho \widehat{C}_i.
\end{equation}
This is the QM-rule for calculation of probabilities of detection.

\section{Coincidence detection}
\label{CAL1}

Coincidence of clicks corresponds to matching of two conditions of   threshold
approaching corresponds to two constraints:
\begin{equation}
\label{BE0c_k}
{\cal E}_1(\tau_1(\omega), \omega)= {\cal E}_d, {\cal E}_2(\tau_2(\omega), \omega)= {\cal E}_d = {\cal E}_d,
\end{equation}
where matching has the form 
\begin{equation}
\label{BE0c_k1}
\tau_{1}(\omega)= \tau_{2}(\omega)= \tau(\omega).
\end{equation}
Our aim is to estimate the probability of coincidence $P_{12}.$ 
To shorter notatin, we set ${\cal E}_i(\omega)\equiv {\cal E}_i(\tau(\omega); \omega)$ or even simply $\Gamma_i, i=1,2.$

We consider the set of random parameters corresponding to coincidence detection:
$A_{12}=\{\omega: {\cal E}_i = {\cal E}_d , i= 1,2 \}.$  $(A_{12}$  is the event 
of coincidence detection).  We have to estimate its probability, $P_{12}= P(A_{12})).$ We shall get a rather rough estimate
which, nevertheless, will be sufficient for our purpose. However, we shall see that better estimates of this probability 
will clarify essentially inter-relation between our ``prequantum classical field theory'',  QM, and experiment. In principle, one may hope 
to derive an approximative expression for $P_{12}$ as we did for probabilities $P_j, j=1,2.$ However, this is a complicated 
probabilistic problem. 

We remark that $A_{12}$ is a subset 
of  the set $$A_{1\times 2}= \{\omega: {\cal E}_1 {\cal E}_2={\cal E}_d^2 \}.$$ Hence, $P(A_{12}) \leq P(A_{1\times 2}).$
And the set  $A_{1\times 2}$ is a subset of the set  
$$
A_{1\times 2 \geq {\cal E}_d^2 } = \{\omega: {\cal E}_1 {\cal E}_2 \geq {\cal E}_d^2 \}
$$ and hence $P(A_{12}) \leq P(A_{1\times 2 \geq {\cal E}_d^2 }).$ The latter probability we can (roughly)  estimate by using 
Chebyshov inequality (which usage is standard for such estimates, cf. \cite{S1}). In the simplest form, for a random variable $u=u(\omega)$ and a  constant $
k>0,$ this inequality has the form: $P(\omega: u \geq k) \leq  \frac{E \vert u(\omega)\vert}{k}.$ In our case
$u= {\cal E}_1 {\cal E}_2, k= {\cal E}_d^2 .$ We have
\begin{equation}
\label{EST}
P(A_{12}) \leq \frac{E {\cal E}_1 {\cal E}_2}{{\cal E}_d^2}.
\end{equation}
We find this average by using the formula of total probability \cite{IP}, \cite{S1}:
\begin{equation}
\label{EST_L}
E {\cal E}_1(\omega) {\cal E}_2(\omega) = \int_0^\infty E 
{\cal E}_1(\tau, \omega) {\cal E}_2(\tau, \omega) P( \tau(\omega)= \tau) d\tau.
\end{equation}
Thus our main problem is to find the correlation of two energies for each instant of time $\tau.$ 
We have 
$$
E{\cal E}_1(\tau, \omega) {\cal E}_2(\tau, \omega)= E w^4(\tau) E \vert \phi_1( \omega)\vert^2 \vert \phi_2( \omega)\vert^2.  
$$
The first factor is known, $E w^4(\tau)= 3 \tau^2.$ 

To find the second factor, we shall use general theory of Gaussian integrals on complex Hilbert space \cite{JMP}, see appendix.
Consider in $H$ ($m$-dimensional complex space)
projection operators $\widehat{A}_k=\vert e_k\rangle \langle e_k\vert, k=1,...,m.$ 
Set $f_{ A_k}(\phi) = \langle \widehat{A}_k \phi, \phi \rangle, \phi \in H,$ the quadratic form corresponding 
to the operator $\widehat{A_k}.$ By (\ref{YY1}), appendix,  we obtain
\begin{equation}
\label{YY1A} E f_{A_1} f_{A_2}  = {\rm Tr} B
\widehat A_1 {\rm Tr}  B \widehat  A_2 + {\rm Tr}  B \widehat  A_2  B
\widehat A_1,
\end{equation}
where $B$ is the covariance operator. We remark that 
\begin{equation}
\label{RHO}
b_{ij}= \rho_{ij} \rm{Tr}B,
\end{equation}
where $\rho=(\rho_{ij})$ is a density operator. So, we consider the prequantum random signal corresponding to 
the  quantum state $\rho$ and we study the problem of detection coincidence for such a signal $\phi(t)\equiv
\phi_\rho(t, \omega).$ 
We have
$${\rm Tr}  B
\widehat  A_i=  b_{ii}, i=1,2.
$$
In the same way
$$
{\rm Tr}  B \widehat  A_2  B
\widehat A_1 =  \vert \langle e_1 \vert B\vert e_2 \rangle\vert^2= \vert b_{12} \vert^2.
$$ 
Finally, we obtain (for the fixed instant of time $\tau)$
$$
E{\cal E}_1(\tau, \omega) {\cal E}_2(\tau, \omega)
= 3 \tau^2  (b_{11} b_{22} + \vert b_{12}\vert^2).
$$
The formula of total probability (\ref{EST_L}) implies
$$
E {\cal E}_1(\tau(\omega); \omega) {\cal E}_2(\tau(\omega); \omega) =
 3 \overline{\tau^2} (b_{11} b_{22} + \vert b_{12}\vert^2) ,
$$
where $\overline{\tau^2}= E \tau^2.$  By (\ref{BE0d2}) we can rewrite this answer in terms of quantum mechanical
density matrix
$$
E {\cal E}_1(\tau(\omega); \omega) {\cal E}_2(\tau(\omega); \omega) 
= 3\overline{\tau^2} \rm{Tr B} (\rho_{ii} \rho_{jj} + \vert \rho_{ij}\vert^2),
$$
Thus by the Chebyshov inequality
\begin{equation}
\label{EST_L1}
P(A_{12}) \leq \frac{3 \rm{Tr} B \overline{\tau^2}}{{\cal E}_d^2} (\rho_{ii} \rho_{jj} + \vert \rho_{ij}\vert^2).
\end{equation}
The quantity 
\begin{equation}
\label{EST_L_K}
\Delta \equiv\sqrt{\overline{\tau^2}}
\end{equation}
 can be interpreted as the time parameter scaling the time intervals between coincidence clicks.
Therefore $\overline{{\cal E}}_\Delta= \Delta \rm{Tr B} $ is average of signal's energy distributed between clicks.
Supppose that 
\begin{equation}
\label{EST_L9}
\epsilon = \frac{\overline{{\cal E}}_\Delta}{{\cal E}_d} <<1.
\end{equation}
In this case 
\begin{equation}
\label{EST_L10}
P(A_{12}) \leq 3 \epsilon^2 (\rho_{ii} \rho_{jj} + \vert \rho_{ij}\vert^2).
\end{equation}
As usual \cite{}, set $g^{(2)}(0)= \frac{P_{12}}{P_1 P_2},$ where $P_1$ and $P_2$ are probabilities of detection in channels $i$ and $j,$ respectively (it is assumed that the later probabilities are positive). We proved, see section \ref{S2}, that 
in TSD model, these probabilities are  equal to quantum probabilities: $P_1= \rho_{11}$ and $P_2 = \rho_{22}.$ 
Hence, 
\begin{equation}
\label{CHCH1}
g^{(2)}(0)= \frac{P_{12}}{P_1 P_2} \leq  3 \epsilon^2 \Big(1 + \frac{\vert \rho_{ij}\vert^2}{\rho_{ii} \rho_{jj}}\Big).
\end{equation}
Since $\epsilon \sim \frac{1}{{\cal E}_d},$ 
by increasing the parameter ${\cal E}_d$ the coefficient $g_{12}(0)$ can be done less than 1.
Hence, our prequantum classical field model can violate (under the selection of a proper threshold) the inequality  $g^{(2)}(0) \geq 1$ which is valid 
for ``standard'' classical signal theory -- its standard version which does not take into account the evident fact that measurement is not simply detection (of the monitoring-type) of continuous classical signals, but creation of discrete counts with the aid of threshold detectors.
Detectors also have to be properly calibrated. 

\section{Appendix: Gaussian integrals}

Details of theory of integration with respect to Gaussian measures on complex Hilbert spaces can be found in \cite{JMP}.

Let $W$ be a real Hilbert space.
Consider  a $\sigma$-additive Gaussian measure $p$ on the
$\sigma$-field of Borel subsets of $W.$ This measure is determined
by its covariance operator $B:  W \to W$ and mean value $m \in W.$
For example, $B$ and $m$ determine the Fourier transform of $p:$
$$
\tilde p(y)= \int_W e^{i(y, \phi)} dp (\phi)=
e^{\frac{1}{2}(By, y) + i(m, y)}, y \in W.
$$
(In probability theory it is called the characteristic functional of the probability distribution $p.)$
In what follows we restrict our considerations to {\it Gaussian
measures with zero mean value}: $ (m,y) = \int_W(y, \psi) d p
(\psi)= 0 $ for any $y \in W.$ Sometimes there will be used the
symbol $p_B$ to denote the Gaussian measure with the covariance
operator $B$ and $m=0.$ We recall that the covariance operator $B$
is defined by its bilinear form
$ (By_1, y_2)=\int (y_1, \phi) (y_2, \phi) dp(\phi),
y_1, y_2 \in W$

Let $Q$ and $P$ be two copies of a real Hilbert space. Let us
consider their Cartesian product $H=Q \times P,$  ``phase space,''
endowed with the symplectic operator $J= \left( \begin{array}{ll}
 0&1\\
-1&0
\end{array}
 \right ).$
Consider the class of Gaussian measures (with zero mean value)
which are invariant with respect to the action of the operator
$J;$ denote this class ${\cal S}(H).$ It is easy to show that $p \in
{\cal S}(H)$ if and only if its covariance operator commutes with the
symplectic operator, \cite{KH1}.
 
As always, we consider complexification of $H$ (which will be
denoted by the same symbol), $H=Q\oplus i P.$ The complex scalar
product is denoted by the symbol $\langle \cdot, \cdot \rangle.$
The space of bounded Hermitian operators acting in $H$ is denoted by
the symbol ${\cal L}_s (H).$

We introduce the complex covariance operator of a measure $p$ on
the complex Hilbert space $H:\;$
$
\langle Dy_1, y_2 \rangle = \int_H \langle y_1, \phi \rangle \langle \phi, y_2 \rangle d p (\phi).
$
Let $p$ be a measure on the Cartesian product $H_1 \times H_2$ of
two complex  Hilbert spaces. Then its covariance operator has the block
structure
\begin{equation}
\label{BLOCK}
D = \left( \begin{array}{ll}
 D_{11} & D_{12}\\
D_{21} & D_{22}\\
 \end{array}
 \right ),
 \end{equation}
where $D_{ii} : H_i \to H_i$ and $D_{ij}: H_j \to H_i.$ The
operator is Hermitian. Hence $D_{ii}^* = D_{ii},$ and $D_{12}^*
= D_{21}.$

Let $H$ be a complex Hilbert space and let $\widehat A \in {\cal
L}_s (H).$ We consider its quadratic form (which will play an important role
in our further considerations)
$
\phi \to f_A(\phi) = \langle \widehat{A} \phi,\phi\rangle.
$
We make a trivial, but ideologically important remark:  $f_A: H \to {\bf R} ,$ is a ``usual
 function'' which is defined point wise.  We use the equality, see, e.g., \cite{KH1}:
\begin{equation}
\label{BL0} \int_H f_A(\phi) dp_D (\phi)={\rm Tr}\; D \widehat A
\end{equation}

\medskip

Let $p$ be a Gaussian measure of the class ${\cal S}(H_1 \times H_2)$ with the
(complex) covariance operator $D$ and let operators $\widehat A_i$ belong to the class ${\cal
L}_s (H_i), i= 1,2.$ Then
\begin{equation}
\label{BL}
\int_{H_1 \times H_2}
f_{A_1}( \phi_1) f_{A_2}(\phi_2) d p (\phi) =
{\rm Tr} D_{11} \widehat A_1 \; {\rm Tr}D_{22} \widehat A_2 +
{\rm Tr}D_{12} \widehat A_2 D_{21} \widehat A_1
\end{equation}

\medskip

This equality   is a consequence of the following general result \cite{KH1}:

 Let $p \in {\cal S}(H)$ with the
(complex) covariance operator $D$ and let $\widehat A_i \in {\cal L}_s (H).$ Then
\begin{equation}
\label{YY1} \int_{H} f_{A_1}( \phi) f_{A_2}(\phi)   dp (\phi), = {\rm Tr} D
\widehat A_1 {\rm Tr} D \widehat A_2 + {\rm Tr} D \widehat A_2 D
\widehat A_1.
\end{equation}

\medskip

{\bf Summary on coincidence probability:} {\it Prequantum classical field model with threshold and properly calibrated
detectors violates predictions of standard classical and semiclassical models, cf. \cite{}, and gives the prediction 
compatible with known experimental data. More detailed experiments of Grangier's type with monitoring of the dependence of 
$g^{(2)}(0)$ on the detection threshold  and source's brightness are on demand.}

\end{document}